\documentclass[a4paper,fleqn]{cas-dc}

\usepackage[numbers]{natbib}

\newcommand{\vecp}{{\bf p}}
\newcommand{\modp}{|{\bf p}|}
\newcommand{\be}{\begin{eqnarray}}
\newcommand{\ee}{\end{eqnarray}}
\newcommand{\nn}{\nonumber \\}

\DeclareMathOperator{\sech}{sech}

\def\hT{{\hat T}}
\def\({\left(}

\def\){\right)}

\begin{document}
\let\WriteBookmarks\relax
\def\floatpagepagefraction{1}
\def\textpagefraction{.001}
\shorttitle{Covariant kinetic theory and transport coefficients for Gribov plasma}
\shortauthors{A Jaiswal et~al.}

\title [mode = title]{Covariant kinetic theory and transport coefficients for Gribov plasma}                      

\author{Amaresh Jaiswal}[orcid=0000-0001-5692-9167]
\ead{a.jaiswal@niser.ac.in}
\author{Najmul Haque}[orcid=0000-0001-6448-089X]
\ead{nhaque@niser.ac.in}
\address{School of Physical Sciences, National Institute of Science Education and Research, HBNI, Jatni 752050, India}


\begin{abstract}
Gribov quantization is a method to improve the infrared dynamics of Yang-Mills theory. We study the thermodynamics and transport properties of a plasma consisting of gluons whose propagator is improved by the Gribov prescription. We first construct thermodynamics of Gribov plasma using the gauge invariant Gribov dispersion relation for interacting gluons. When the Gribov parameter in the dispersion relation is temperature dependent, one expects a mean field correction to the Boltzmann equation. We formulate covariant kinetic theory for the Gribov plasma and determine the mean-field contribution in the Boltzmann equation. This leads to a quasiparticle like framework with a bag correction to pressure and energy density which mimics confinement. The temperature dependence of the Gribov parameter and bag pressure is fixed by matching with lattice results for a system of gluons. Finally we calculate the temperature dependence of the transport coefficients, i.e., bulk and shear viscosities.
\end{abstract}


\begin{keywords}
Relativistic heavy-ion collisions\sep Hydrodynamic models\sep Relativistic fluid dynamics\sep Quantum chromodynamics
\end{keywords}


\maketitle

\bigskip

\section{Introduction}

At extremely high temperature and/or density, the fundamental constituents of hadrons, i.e., quarks and gluons, are expected to become the active degrees of freedom whose interactions are governed by quantum chromodynamics (QCD). In high energy heavy-ion collisions, where extremely high values of temperature are achieved, the quarks and gluons inside the nucleons are liberated over a relatively large volume forming a quark-gluon plasma (QGP). Contrary to the naive early expectations that QGP is a weakly interacting gas, the experimental data from relativistic heavy-ion collisions \cite{ALICE:2011ab, Aamodt:2011by, Adare:2011tg, Chatrchyan:2012wg, Aad:2013xma, Adamczyk:2013waa} provided evidence for it to be a strongly interacting and correlated system \cite{Gyulassy:2004zy, Shuryak:2004cy}. This sparked the interest to study relativistic heavy-ion collisions within the framework of dissipative hydrodynamics and to estimate various transport coefficients of the QCD system \cite{Romatschke:2007mq, Romatschke:2009kr, Song:2010mg, Jeon:2015dfa}; see also Refs.~\cite{Heinz:2013th, Jaiswal:2016hex, Florkowski:2017olj} for recent reviews.

At very high temperatures, QCD thermodynamics is well established within the framework of resummed perturbation theory \cite{Andersen:2004fp, Andersen:2010ct, Andersen:2011sf, Haque:2013sja, Haque:2014rua, Andersen:2015eoa} as well as from lattice calculations \cite{Borsanyi:2013bia, Bazavov:2014pvz}. On the other hand, understanding the transport properties of QCD still remains a challenge \cite{Meyer:2011gj}. For instance, the coefficients of shear and bulk viscosities have been calculated in the high temperature regime using perturbation theory \cite{Arnold:2000dr, Arnold:2003zc, Arnold:2006fz, Moore:2008ws}. However, in the phenomenologically relevant temperature regime, i.e., near critical temperature, results from lattice calculations are plagued by large uncertainties and therefore still inconclusive \cite{Meyer:2007dy, Meyer:2007ic}. The estimation of transport coefficients of QCD matter is very important for relativistic heavy-ion phenomenology and is currently under intense investigation.

Due to long range correlations in the system leading to confinement of colored degrees of freedom, the infrared regime of QCD is strongly coupled. This feature is purely non-perturbative and therefore beyond the scope of conventional perturbative techniques \cite{Andersen:2004fp}. An efficient way to treat the infrared regime of QCD is to consider Gribov-Zwanziger prescription \cite{Gribov:1977wm, Zwanziger:1989mf}. This method improves the infrared dynamics of Yang-Mills theory by fixing residual gauge transformations that remains after employing the Faddeev-Popov quantization. The Gribov-Zwanziger prescription leads to an infrared-improved dispersion relations for gluons~\cite{Gribov:1977wm}. As a consequence, a new energy scale, the Gribov parameter $\gamma_G$, is introduced which explicitly breaks the conformal symmetry of the theory.

The Gribov-Zwanziger approach has recently attracted a lot of attention in the heavy-ion theory community after it was generalized to finite temperature \cite{Zwanziger:2004np, Zwanziger:2006sc}. It was shown that $\gamma_G$, being an intrinsic Yang-Mills scale, significantly improves the infrared behavior of QCD and leads to good agreement with lattice results for thermodynamic quantities \cite{Fukushima:2013xsa}. The Gribov dispersion relation provides a straightforward way to incorporate the effects of residual confinement on the transport properties of QGP. In the context of kinetic theory and hydrodynamics, it was employed for the first time in Refs.~\cite{Florkowski:2015dmm, Florkowski:2015rua, Begun:2016lgx} in a boost-invariant setup. The effect on observables like dilepton rate and quark number susceptibility has also been examined lately \cite{Bandyopadhyay:2015wua}. However, to the best of our knowledge, a covariant kinetic theory for Gribov plasma has not yet been formulated.

In this paper, we study the thermodynamics and transport properties of a plasma consisting of gluons whose propagator is improved by the Gribov prescription. We first construct thermodynamics of Gribov plasma using the gauge invariant Gribov dispersion relation for interacting gluons. Further, we formulate, for the first time, a covariant kinetic theory for the Gribov plasma and determine the mean-field contribution in the Boltzmann equation by considering the Gribov parameter in the dispersion relation to be temperature dependent. This leads to a quasiparticle like framework with a bag correction to pressure and energy density. The temperature dependence of the Gribov parameter and bag pressure is fixed by matching with lattice results for a system of gluons. Finally we calculate the temperature dependence of the transport coefficients. Throughout the paper we use natural units with $c=\hbar=k_B=1$. We denote three-vectors by bold font and four-vectors in standard font. The center-dot represents scalar product of four-vectors with the metric \mbox{$g^{\mu\nu}={\rm diag}(+1,-1,-1,-1)$}.

\medskip

\section{Thermodynamics of Gribov plasma}

The gluon propagator with Gribov term is \cite{Gribov:1977wm}
\begin{equation}\label{propagator}
D^{\mu \nu}(p)=\left[\delta^{\mu\nu}-(1-\xi) \frac{p^\mu p^\nu}{p^{2}}\right] \frac{p^2}{p^4+\gamma_G^4} \, ,
\end{equation}
where $p^\mu$ is the gluon four-momentum, $\xi$ is the gauge parameter, $p^2=p\cdot\! p = p^\mu p_\mu$ and $\gamma_G$ is the Gribov parameter. The propagator in the above equation has quasiparticle poles at $E_\pm=\sqrt{\modp^2 \pm i\gamma_G^2}$, where $\vecp$ is the three-momentum of the gluons. Although the  poles of the gluon propagator are shifted to an unphysical location, $p^2=\pm i\gamma_G^2$, it was shown that the glueball channel contains a physical cut having positive spectral function \cite{Zwanziger:1989mf}. While the Gribov parameter was originally introduced by Gribov to explain the low energy confinement, we note that the study of deconfined nuclear matter with Gribov parametrization is not new and has been explored extensively in the literature~\cite{Zwanziger:2004np, Zwanziger:2006sc, Fukushima:2013xsa, Florkowski:2015dmm, Florkowski:2015rua, Begun:2016lgx, Bandyopadhyay:2015wua, Su:2014rma}.

The justifications for employing Gribov prescription to describe deconfined nuclear matter is well known and can be found in literature, for instance in Ref.~\cite{Zwanziger:2006sc}. It has been explained in details in Ref.~\cite{Zwanziger:2006sc} that the evidence for confinement was observed from lattice simulations where it was found that the long-distance behavior of the color-Coulomb potential grows linearly with distance $R$ as $V_{\rm coul}(R) \sim \sigma_{\text {coul}}\, R$ with $\sigma_{\text {coul}}\sim 3\sigma$ and $\sigma$ being the physical string tension between a pair of external quarks. It was also found numerically that the long-distance behavior of $V_{\text {coul }}(R)$ is consistent with a linear increase, $\sigma_{\text {coul }}>0,$ above the phase transition temperature, $T>T_c$, where $\sigma$ vanishes. Investigation of the temperature dependence of $\sigma_{\text {coul}}$ revealed that in the deconfined phase, the Coulomb string tension increases with $T$, which is consistent with a magnetic mass $\sigma_{\rm coul}^{1/2}(T) \sim g_s^{2}(T)\,T$, with $g_s$ being the strong coupling~\cite{Nakagawa:2006fk}. Thus, from the numerical evidence and also from the explanation provided in Ref.~\cite{Zwanziger:2006sc}, one can say that the Gribov parametrization works well in the confined phase of QCD as well as in the deconfined phase.

The Gribov parameter $\gamma_G$ in Eq.~\eqref{propagator} can be obtained by self-consistently solving a gap equation which is defined to infinite loop orders \cite{Gribov:1977wm, Zwanziger:1989mf}. This is, in general, a non trivial task. At one-loop order and at asymptotically high temperatures, one obtains $\gamma_G\sim g_s^2\,T$~\cite{Fukushima:2013xsa}. At finite temperature effective gluon mass appears as electric scale ($g_s\,T$) in the theory whereas Gribov parameter behaves like magnetic scale ($g_s^2\, T$) at high temperature. Therefore, in principle Gribov term can be introduced along with the thermal gluon mass to calculate various properties of deconfined matter in a perturbative way, as done in Refs.~\cite{Bandyopadhyay:2015wua, Su:2014rma}. However, in the current calculation we do not consider perturbative form of the Gribov parameter. In this work, we consider $\gamma_G(T)$ to be a general function of temperature whose functional form is determined by fitting the lattice QCD equation of state. Therefore near cross-over region, where perturbative calculation fails and we can't use asymptotic forms of thermal quark mass or Gribov parameter, the present analysis will be useful. However, in such quasiparticle frameworks, it is important to ensure that the thermodynamic consistency is maintained.

In order to ensure thermodynamic consistency at equilibrium, one can introduce an additional effective mean field through a bag function $B_0(T)$ \cite{Gorenstein:1995vm}. In a Lorentz covariant way, this can be achieved by modifying the definition of the equilibrium energy-momentum tensor \cite{Chakraborty:2010fr, Romatschke:2011qp, Albright:2015fpa, Tinti:2016bav,Jeon:1994if, Jeon:1995zm}
\begin{equation}\label{Tmunu_eq}
T_{(0)}^{\mu\nu} = \int dp \, p^\mu p^\nu \, f_0  \, + B_0(T) \, g^{\mu\nu},
\end{equation}
where $g^{\mu\nu}=\rm{diag}(1,-1,-1,-1)$ is the metric tensor and $B_0(T)$ is the bag function which can be determined by requiring thermodynamic consistency in equilibrium. Here $f_0$ is the equilibrium distribution function which can have contribution from two quasiparticle energy poles, $E_\pm$, of the gluon propagator. In the above equation, $dp$ is the Lorentz invariant momentum integration measure defined as
\begin{equation}\label{int_mes}
\int dp \equiv  \frac{g}{(2\pi)^4}\! \int\! d^4 p \; 2\Theta\!\(p^0\) (2\pi) \, \delta\!\(p^2+\frac{\gamma_G^4}{p^2}\),
\end{equation}
where $g$ is the degeneracy factor and $\Theta$ is the Heaviside step function to ensure that only the energy states which has positive real part are considered.

The equilibrium pressure and energy density can be obtained from Eq.~\eqref{Tmunu_eq} using the definitions
\begin{align}
P_0 \, &\equiv -\frac{1}{3}\Delta_{\mu\nu} T_{(0)}^{\mu\nu} = P_G - B_0, \label{prs}\\
\varepsilon_0 \, &\equiv\, u_\mu u_\nu T_{(0)}^{\mu\nu}\, =\, \varepsilon_G + B_0, \label{eng_den}
\end{align}
where $u^\mu$ is the fluid four-velocity satisfying $u_\mu u^\mu=1$ and $\Delta^{\mu\nu}\equiv g^{\mu\nu}-u^\mu u^\nu$. In the above equation, $P_G$ and $\varepsilon_G$ are the particle contribution to pressure and energy density, respectively, arising from the distribution function of the equilibrium Gribov plasma. In the local rest frame, i.e., $u^\mu=(1,0,0,0)$, $P_G$ and $\varepsilon_G$ are given by
\begin{align}
P_G &= \frac{g }{(2 \pi)^{3}}\int \mathrm{d}^{3} \vecp\ \frac{\modp^2}{6}\(\frac{f_0^+}{E_+} +\frac{f_0^-}{E_-}\), \label{prs_G} \\
\varepsilon_G &= \frac{g}{(2 \pi)^3}\int \mathrm{d}^{3} \vecp\ \frac{1}{2} \left( f_0^+ \, E_+ + f_0^- \, E_- \right), \label{eng_den_G}
\end{align}
where the form of the equilibrium distribution function for the Gribov plasma at the two energy poles are
\begin{equation}\label{eq_dist}
f_0^\pm = \left[  \exp\!\(\frac{E_\pm}{T}\) -1 \right]^{-1}.
\end{equation}
The momentum integrations in Eqs.~\eqref{prs_G} and~\eqref{eng_den_G} can be done analytically for the distribution function given in the above equation.

After performing the integrals in Eqs.~\eqref{prs_G} and \eqref{eng_den_G}, we obtain
\begin{align}
P_G &= \frac{gz^2 T^4}{4\pi^2}\sum_{l=1}^{\infty}\frac{1}{l^2}\left[ iK_2\(\sqrt{i}\,lz\) + i \to -i \right], \label{prs_K}\\
\varepsilon_G &= \frac{gz^3 T^4}{4\pi^2}\!\sum_{l=1}^{\infty}\frac{1}{l}\!\left[ i^{3/2}K_3\!\(\!\sqrt{i}\,lz\) + i\! \to\! -i\right] - P_G, \label{eng_den_K}
\end{align}
where $i=\sqrt{-1}$ is the unit imaginary number, $K_n(x)$ are the modified Bessel functions of second kind of order $n$ and $z\equiv\gamma_G/T$. It is important to note that the terms in square brackets in the above equations are real since it is just sum of two complex conjugates. Therefore the thermodynamics of Gribov plasma is real even though the Gribov energy-momentum dispersion relations, given by the poles of the propagator in Eq.~\eqref{propagator}, are complex. In the calculations, we employ the identity, $\sqrt{\pm i}=(1\pm i)/\sqrt{2}$.

In order to calculate the particle contribution to entropy density of Gribov plasma, we use the Boltzmann definition of entropy density for bosons,
\begin{equation}\label{entropy_bolt}
s_G = -\int dp\, (u\!\cdot\!p) \left[f_0 \ln f_0 - (1+f_0) \ln (1+f_0)\right].
\end{equation}
Again going to the local rest frame as well as using Eqs.~\eqref{prs_G} and \eqref{eng_den_G}, we find that $s_G=(P_G+\varepsilon_G)/T$. On the other hand, from Eqs.~\eqref{prs} and \eqref{eng_den}, we see that $s_G=(P_0+\varepsilon_0)/T=s_0$ implying that the mean field contributions through $B_0(T)$ is absent in equilibrium entropy density. It is important to note that for large values of $\gamma_G/T$, the above expression for entropy density might result in negative values. On the other hand, Coulomb gauge calculations for the Gribov-Zwanziger plasma leads to positive entropy density for all values of $\gamma_G/T$ \cite{Begun:2016lgx}. However, as demonstrated in Sec.~\ref{sec5}, the value of $\gamma_G/T$ is fixed by matching the above expression for entropy density with that obtained from lattice QCD calculations for gluonic plasma. As the lattice results for entropy density is positive, therefore we also have the same by construction. For thermodynamic consistency in equilibrium, we require that $s_0 = dP_0/dT$. Taking derivative of Eq.~\eqref{prs_G} with respect to $T$ and rearranging, we find that the thermodynamic consistency is maintained only if 
\begin{equation}\label{th_cons}
\frac{dB_0}{dT} + \frac{g}{(2\pi)^3}\gamma_G\dfrac{d\gamma_G}{dT}\int\mathrm{d}^{3} \vecp\, \frac{i}{2}\! \(\frac{f_0^+}{E_+} -\frac{f_0^-}{E_-}\) = 0.
\end{equation}
The above integral can be done analytically to obtain
\begin{equation}\label{th_cons_K}
\frac{dB_0}{dT} + \frac{gz^3 T^3\kappa}{4\pi^2}\!\sum_{l=1}^{\infty}\!\frac{1}{l}\!\left[ i^{3/2}K_1\!\!\(\!\sqrt{i}\,lz\!\) + i \to -i\right]\! = 0,
\end{equation}
where $\kappa\equiv\frac{T}{\gamma_G}\frac{d\gamma_G}{dT}$. We again note that the expression in square brackets in the above equation is real.

\medskip

\section{Out-of-equilibrium and the Boltzmann equation}

For the general non-equilibrium case, one can consider the energy momentum tensor $T^{\mu\nu}=T^{\mu\nu}_{(0)}+\Delta T^{\mu\nu}$. In terms of the distribution function, we can write
\begin{equation}\label{Tmunu}
T^{\mu\nu} = \int dp \, p^\mu p^\nu \, f  \, + B(T)\,g^{\mu\nu},
\end{equation}
where $f$ is the non-equilibrium distribution function and $B=B_0+\Delta B$ is the nonequilibrium bag pressure. In the above equation, the four-momentum of the gluons now have energy poles at $E_\pm=\sqrt{\modp^2 \pm i\Gamma_G^2}$ where we denote the out-of-equilibrium Gribov parameter as $\Gamma_G^2=\gamma_G^2+\Delta\gamma_G^2$. Since we consider a system without any conserved charges, we use the Landau frame definition for fluid velocity, $u_\mu T^{\mu\nu}=\varepsilon u^\nu$, which is the most natural choice. In the local rest frame, this leads to the matching conditions $\Delta T^{00}=0$ and $\Delta T^{0i}=0$.

The energy-momentum conservation of the system requires that the four-divergence of the energy-momentum tensor vanishes, i.e., $\partial_\mu T^{\mu\nu}=0$. The four-divergence of $T^{\mu\nu}$ in Eq.~\eqref{Tmunu} leads to
\begin{align}\label{dTmunu}
&\partial^\nu B + \frac{\partial^\nu\Gamma_G}{\Gamma_G}\int dp\,p^2\,f \nn
&+ \int dp\, p^\nu \left[ p^\mu\partial_\mu f + \frac{1}{\Gamma_G}\(\partial_\mu\Gamma_G\) \partial^\mu_{(p)}\!\(p^2f\) \right] = 0,
\end{align}
where $\partial^\mu_{(p)}$ represents derivative with respect to four-momenta. To obtain the above result, we used the trick
\begin{equation}\label{trick}
p^\mu\partial_\mu \delta\!\(\!p^2+\frac{\Gamma_G^4}{p^2}\!\) = -\frac{p^2}{\Gamma_G}\(\partial_\mu\Gamma_G\) \partial^\mu_{(p)}\delta\!\(\!p^2+\frac{\Gamma_G^4}{p^2}\!\)\!,
\end{equation}
and subsequently performed integration by parts. After simplifying the integral in the second term of Eq.~\eqref{dTmunu}, we find that the first two terms in Eq.~\eqref{dTmunu} is exactly of the same form as that required from thermodynamic consistency, i.e., the left hand side of Eq.~\eqref{th_cons}, but with equilibrium quantities $B_0$, $\gamma_G$ and $f_0$ replaced by out-of-equilibrium quantities $B$, $\Gamma_G$ and $f$.

We extend the thermodynamic consistency condition, Eq.~\eqref{th_cons}, to hold out of equilibrium which leads to the Boltzmann equation 
\begin{equation}\label{Boltz_eq}
p^\mu\partial_\mu f + \frac{p^2}{\Gamma_G}\(\partial_\mu\Gamma_G\) \partial^\mu_{(p)}f = C[f],
\end{equation}
where $C[f]$ is the collision kernel and the term containing derivative of $\Gamma_G$ is the mean field term. Indeed the concept of extending the thermodynamic relations to hold out of equilibrium is quite powerful and has been previously applied in the context of medium dependent masses of classical Boltzmann particles~\cite{Romatschke:2011qp, Czajka:2017wdo}. In arriving at the above equation, one has to keep in mind that the mean field force, $F^\alpha \sim \partial^\alpha\Gamma_G$, is orthogonal to gluon momentum, i.e., $p_\alpha F^\alpha=0$. This restriction on the mean field force is imposed to satisfy the second matching condition $\Delta T^{0i}=0$ implying that $\delta f$ (which we compute later) can not have a vector component. Equation~\eqref{Boltz_eq} represents the main result of present work. Note that, in order to satisfy the conservation of energy-momentum tensor, the first moment of collision kernel must vanish, i.e., $\int dp\, p^\nu\, C[f]=0$. In the present work, we assume relaxation-time approximation for the collision term \cite{Anderson:1974},
\begin{equation}\label{RTA}
C[f] = -\frac{\(u\!\cdot\! p\)}{\tau_R}\,\Delta f,
\end{equation}
where $\tau_R$ is the relaxation time which is independent of particle momentum and $\Delta f\equiv f-f_0$.

In order to calculate the nonequilibrium corrections to the Gribov parameter, $\Delta\gamma_G^2$, the distribution function, $\Delta f$, and the bag pressure, $\Delta B$, we follow the prescription described in Ref.~\cite{Czajka:2017wdo}. The out-of-equilibrium phase-space density can be written as $f=F_0+\delta f$ where $\delta f$ is the deviation from equilibrium and $F_0$ has the form of local equilibrium distribution but with nonequilibrium Gribov parameter $\Gamma_G$ in the energy dispersion relation for gluons. Extending the arguments of Refs.~\cite{Jeon:1994if, Jeon:1995zm} for Gribov plasma and following the analysis in Ref.~\cite{Czajka:2017wdo}, we obtain
\begin{align}
\Delta\gamma_G^2 =&~ \frac{2\,\kappa\,T}{J_{10}}\int\! dp\,p^2\,\delta f, \label{Deltag}\\
\Delta f =&~ \delta f -  \frac{f_0\,\tilde{f_0}\,\kappa}{J_{10}\,(u\!\cdot\!p)} \int\! dp\,p^2\,\delta f, \label{Deltaf}\\
\Delta B =&~-\frac{I_{00}\,\kappa\,T}{J_{10}}\int\! dp\,p^2\,\delta f, \label{DeltaB}
\end{align}
where $\tilde{f_0}\equiv 1+f_0$ for bosons and the right-hand-side of above equations are evaluated with equilibrium $\gamma_G$ in the energy dispersion relation for gluons. In the following we will always consider the gluon momentum equilibrium $\gamma_G$ in the energy dispersion relation. The integral coefficients appearing in Eqs.~\eqref{Deltag}-\eqref{DeltaB}, are defined as
\begin{equation}\label{Inq}
I_{nq} = \frac{1}{(2q+1)!!}\int\! dp\, (u\!\cdot\!p)^{n-2q} \(-\Delta_{\mu\nu}p^\mu p^\nu\)^q f_0,
\end{equation}
and similarly for $J_{nq}$ but with $f_0$ replaced by $f_0\tilde{f_0}$. The particular coefficients appearing in Eqs.~\eqref{Deltag}-\eqref{DeltaB} can be evaluated as
\begin{align}
I_{00} =&\, \frac{g\, T^2 z }{4\,\pi^2} \sum_{l=1}^{\infty}\Big[\sqrt{i}\, K_1(\sqrt{i}\,lz) + i \to -i \Big], \label{I10}\\
J_{10} =&\, \frac{g\, T^3 z^2 }{4\,\pi^2} \sum_{l=1}^{\infty}\Big[i\, K_2(\sqrt{i}\,lz) + i \to -i \Big]. \label{J10}
\end{align}
Note that Eqs.~\eqref{Boltz_eq}, \eqref{RTA} and~\eqref{Deltaf} has to be solved self consistently to evaluate $\delta f$ which is required for calculation of the transport coefficients.

The matching condition, $\Delta T^{00}=\int dp\, (u\cdot p)^2 \Delta f=0$, along with Eq.~\eqref{Deltaf} leads to
\begin{equation}\label{matcon}
\int dp \left[ (u\cdot p)^2 - \kappa\, p^2 \right] \delta f = 0.
\end{equation}
We also note that the sound velocity, $c_s^2\equiv dP_0/d\varepsilon_0$, can be obtained as
\begin{equation}\label{cs2gri}
\frac{1}{3}-c_s^2 = \frac{(1-\kappa)\,\bar{J}_{10}}{3\(J_{30}-\kappa\,\bar{J}_{10}\),}
\end{equation}
where we have defined $\bar{J}_{nq}$ similar to $I_{nq}$ in Eq.~\eqref{Inq} but with $f_0$ replaced by $p^2f_0\tilde{f_0}$. The integral coefficients appearing in the above equation are given by
\begin{align}
\bar{J}_{10} =&\, -\frac{g\, T^5 z^4 }{4\,\pi^2} \sum_{l=1}^{\infty}\Big[ K_2(\sqrt{i}\,lz) + i \to -i \Big], \label{Jb10}\\
J_{30} =&\, -\frac{g\, T^5 z^4 }{8\,\pi^2} \sum_{l=1}^{\infty} \Big[ \left( K_4 + K_2 \right) + i \to -i\Big], \label{J30}
\end{align}
where, if not explicitly mentioned, the argument of the special functions are implicitly taken as $(\sqrt{i}\,lz)$.

\medskip

\section{Transport coefficients}

The dissipative quantities can be written in terms of $\delta f$ as,
\begin{equation}\label{dissi_del}
\pi^{\mu\nu} = \Delta^{\mu\nu}_{\alpha\beta} \!\int\! dp\,p^\alpha p^\beta \delta f,~~
\Pi = -\frac{\Delta_{\alpha\beta}}{3}\!\int\! dp\,p^\alpha p^\beta \delta f,
\end{equation}
where $\pi^{\mu\nu}$ is the shear stress tensor, $\Pi$ is the bulk viscous pressure and $\Delta^{\mu\nu}_{\alpha\beta}\equiv\frac{1}{2} (\Delta^{\mu}_{\alpha}\Delta^{\nu}_{\beta} + \Delta^{\mu}_{\beta}\Delta^{\nu}_{\alpha} - \frac{2}{3} \Delta^{\mu\nu}\Delta_{\alpha\beta})$ is the symmetric traceless projector orthogonal to $u^\mu$. To obtain the transport coefficients, we calculate first-order derivative correction to the distribution function from the Boltzmann equation in the relaxation-time approximation, Eqs.~\eqref{Boltz_eq} and~\eqref{RTA}, using a Chapman-Enskog like iterative solution \cite{Jaiswal:2013npa, Bhalerao:2013pza}. The first-order correction is given by
\begin{equation}\label{deltaf1}
\Delta f_1 = -\frac{\tau_R}{u\!\cdot\!p}\left[ p^\mu\partial_\mu f_0 + \frac{p^2}{\gamma_G}\(\partial_\mu\gamma_G\) \partial^\mu_{(p)}f_0 \right].
\end{equation}
From the above equation, we see that the derivative of the equilibrium distribution function, Eq.~\eqref{eq_dist}, would lead to temperature derivatives. Using thermodynamic identities and projection of energy-momentum conservation equation, $\partial_\mu T^{\mu\nu}=0$, along and orthogonal to the fluid velocity, we obtain 
\begin{equation}\label{temp_deriv}
\frac{\dot T}{T} = -c_s^2\,\theta + {\cal O}(\partial^2), \quad 
\frac{\nabla^\mu T}{T} = \dot u^\mu + {\cal O}(\partial^2),
\end{equation}
where $\dot A\equiv u^\mu\partial_\mu A$ is the comoving derivative, $\nabla^\mu A \equiv \Delta^{\mu\nu}\partial_\nu A$ is derivative projected orthogonal to fluid velocity and $\theta\equiv\partial_\mu u^\mu$ is the expansion scalar.

Using identities of Eq.~\eqref{temp_deriv} in Eq.~\eqref{deltaf1}, we get
\begin{align}\label{Deltaf1}
\Delta f_1 =&\, \frac{f_0\,\tilde{f_0}\,\tau_R}{T\(u\!\cdot\!p\)}\!\left[ \(\frac{1}{3} - \kappa\,c_s^2\)\!p^2 - \(\frac{1}{3} - c_s^2\)\! \(u\!\cdot\!p\)^2 \right]\!\theta \nn
&\qquad\qquad  + \frac{f_0\,\tilde{f_0}\,\tau_R}{T\(u\!\cdot\!p\)}\,p^\mu\, p^\nu\, \sigma_{\mu\nu}\,,
\end{align}
where $\sigma^{\mu\nu}\equiv\Delta^{\mu\nu}_{\alpha\beta} (\nabla^\alpha u^\beta)$ is the velocity stress tensor. Comparing the above equation with Eq.~\eqref{Deltaf}, we get
\begin{align}\label{deltaf_1}
\!\!\!\!\!\!\!\!\!\!\delta f_1 =&\, -\frac{f_0\,\tilde{f_0}\,\tau_R}{T\(u\!\cdot\!p\)}\(\frac{1}{3} - c_s^2\)\!\left[ \(u\!\cdot\!p\)^2 - \frac{J_{30}-\kappa\,\bar{J}_{10}}{J_{10}-\kappa\,\bar{J}_{{\textrm -}10}} \right]\!\theta \nn
&  + \frac{f_0\,\tilde{f_0}\,\tau_R}{T\(u\!\cdot\!p\)}\,p^\mu\, p^\nu\, \sigma_{\mu\nu}\,,
\end{align}
where the unknown integral coefficient $\bar{J}_{{\textrm -}10}$ can be obtained in terms of $K_1(\sqrt{i}\,lz)$ and the Bickley function denoted by $\widetilde{K}_1(\sqrt{i}\,lz)\equiv\int_0^\infty d\phi\,\sech\phi\,\exp(-\sqrt{i}\,lz\cosh\phi)$,
\begin{equation}\label{Jbm10}
\bar{J}_{{\textrm -}10} = \frac{g\, T^3 z^3 }{4\,\pi^2} \sum_{l=1}^{\infty}l\Big[i^{3/2}\left(K_1 - \widetilde{K}_1\right) + i \to -i\Big]. 
\end{equation}
We will use Eq.~\eqref{deltaf_1} in Eq.~\eqref{dissi_del} to obtain the transport coefficients.

Substituting the expression for $\delta f_1$ from Eq.~\eqref{deltaf_1} in Eq.~\eqref{dissi_del}, we obtain the Navier-Stokes equation for shear stress tensor,
\begin{equation}\label{shear}
\pi^{\mu\nu} = 2\,\eta\,\sigma^{\mu\nu}.
\end{equation}
Here the coefficient of shear viscosity is obtained in terms of the integral coefficient, $\eta=\tau_RJ_{32}/T$, and is given by
\begin{align}\label{eta}
\!\!\!\!\!\!\!\!\!\!\eta = - \frac{g\,\tau_R\, T^4 z^5 }{960\,\pi^2} \sum_{l=1}^{\infty}l\Big[&\sqrt{i}\left( K_5 - 7K_3 + 22K_1 - 16\widetilde{K}_1 \right) \nn
&+ i \to -i\Big],
\end{align}
In order to compute the coefficient of bulk viscosity, we first use the matching condition, Eq.~\eqref{matcon} in Eq.~\eqref{dissi_del} to rewrite the bulk viscous pressure as 
\begin{equation}\label{bulk_def}
\Pi = -\frac{1}{3}\(1-\kappa\)\int dp\, p^2\, \delta f.
\end{equation}
Substituting $\delta f_1$ from Eq.~\eqref{deltaf_1} in the above equation, we get the Navier-Stokes equation for bulk viscous pressure
\begin{equation}\label{bulk}
\Pi=-\,\zeta\,\theta.
\end{equation}
Here we obtain the coefficients of bulk viscosity $\zeta$ as
\begin{equation}
\zeta = \frac{\tau_R\,\gamma_G^4\,(1-\kappa)^2}{9\,T} \left[\! \frac{J_{10}\,J_{10}}{J_{30}-\kappa \bar{J}_{10}} - \frac{J_{{\textrm -}10}\,J_{10}}{J_{10}-\kappa \bar{J}_{{\textrm -}10}} \!\right] \label{zeta},
\end{equation}
where $\displaystyle{J_{{\textrm -}10} = \frac{g\, T z }{4\,\pi^2} \sum_{l=1}^{\infty}l\Big[\sqrt{i}\left(K_1 - \widetilde{K}_1\right) + i \to -i\Big].}$


\section{Results and discussions}\label{sec5}

In order to calculate the transport coefficients of the Gribov plasma, we first fix the equilibrium thermodynamic quantities by matching the temperature dependence of the scaled trace anomaly of lattice results \cite{Borsanyi:2012ve}. For numerical convenience, we use analytic fit for the trace anomaly
\begin{eqnarray}\label{trace} 
\frac{{\cal I}^{\rm fit}_\text{0}(T) }{T^4}  &=& \exp\!\Big[\!-\Big(\,\!\frac{h_1}{\hT}+\frac{h_2}{\hT^2}\Big)\Big] \\
&&\hspace{-0.5cm}\times\! \left[\frac{h_0}{1+h_3 \hT^2}+\frac{f_0\big[\tanh(f_1 \hT+f_2)+1\big]}{1+g_1 \hT+g_2 \hT^2}\right] ,\nonumber
\end{eqnarray}
with $\hT\equiv T/T_c$. The above functional form for the fit was taken from Ref.~\cite{Borsanyi:2010cj}. The fit parameters were obtained as: $h_0=0.234236$, $h_1=-1.83115$, $h_2=2.92255$, $h_3=0.0703023$, $f_0=0.328193$, $f_1=62.3957$, $f_2=-62.5558$, $g_1=-1.98532$ and $g_2=1.08707$. Using the above equation, the equilibrium pressure is obtained by
\begin{equation}\label{P_func}
\frac{P_0(T)}{T^4}=\frac{P_0(T_0)}{T_0^4}+\int_{T_0}^T \frac{dT'}{T'}\frac{{\cal I}^{\rm fit}_\text{0}(T')}{T'^4} \, ,
\end{equation}
where $P_0(T_0=0.7\,T_c)$ is again taken from Ref.~\cite{Borsanyi:2012ve}. In order to fix rest of the thermodynamic quantities, we employed the fact that $s_0=dP_0/dT=(\varepsilon_0+P_0)/T=(\varepsilon_G+P_G)/T$ and used Eqs.~\eqref{prs}, \eqref{eng_den}, \eqref{prs_K} and \eqref{eng_den_K}.

\begin{figure}[t!]
\center
\includegraphics[width=\linewidth]{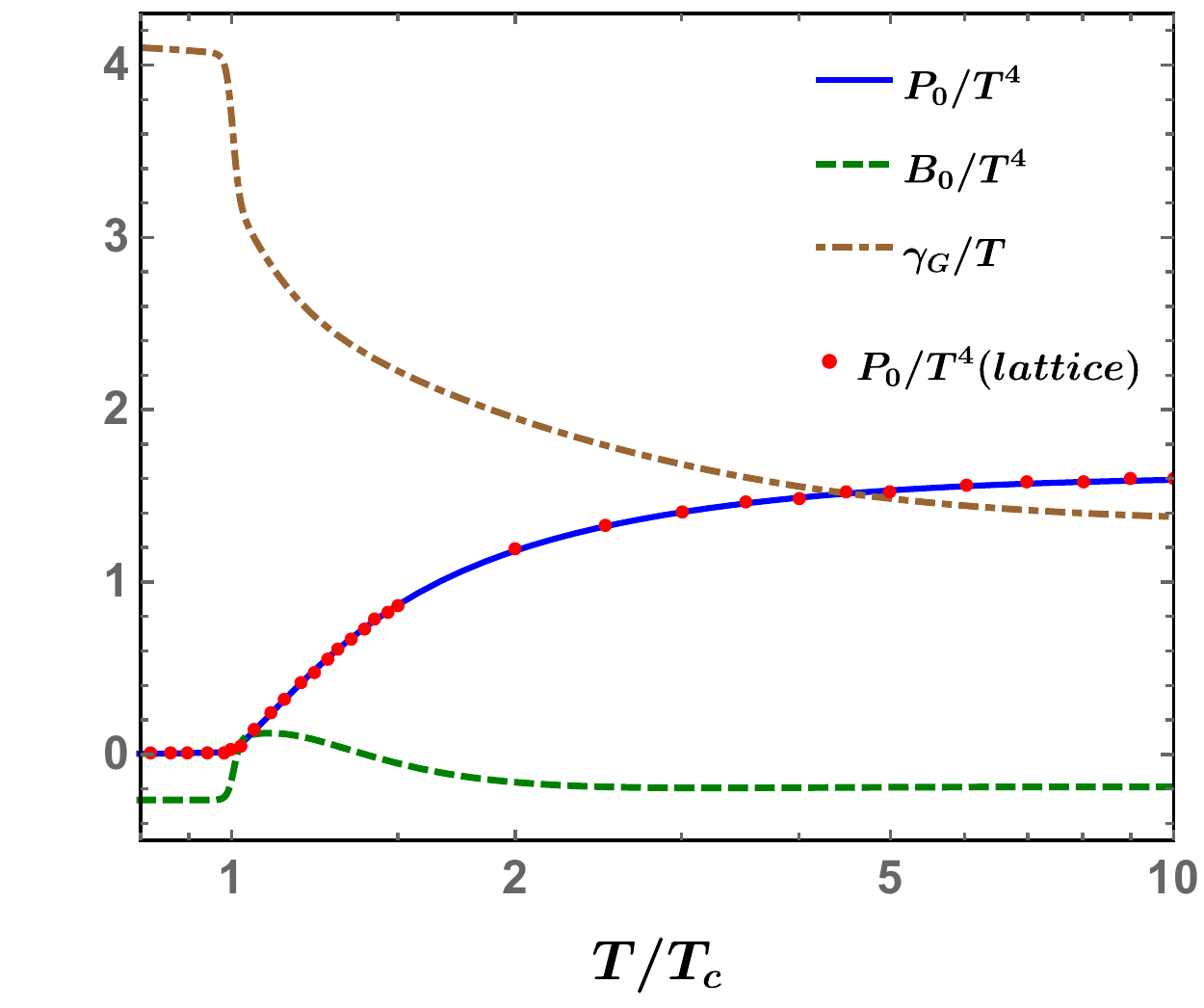}
\caption{Temperature dependence of various thermodynamic quantities obtained after matching with lattice data.}
\label{thermo}
\end{figure} 

In Fig.~\ref{thermo}, we show temperature dependence of various thermodynamic quantities obtained after matching with lattice data. We see that the fitted value of the pressure (blue solid line) matches with the corresponding lattice result for pressure (red dots) at the level of $1-2\%$ accuracy. We also see that the scaled Gribov parameter, $\gamma_G/T$, (brown dashed-dotted line) decreases with temperature and tends to saturate to a constant value at very high temperatures. Since the introduction of $\gamma_G$ results in explicit breaking of the conformal symmetry, this feature has important implications for bulk viscosity as discussed later. Note that because of the choice of covariant gauge in the present calculation, the value of the Gribov parameter near $T_c$ that we extracted is nearly half compared to that obtained in Ref.~\cite{Begun:2016lgx} where the authors considered Coulomb gauge. At very high temperature, we find the qualitative behavior $\gamma_G\sim T$ which is consistent with perturbative QCD prediction $g_s^2T$~\cite{Fukushima:2013xsa} as the temperature dependence of $g_s$ becomes weak. The scaled bag parameter (green dashed line) has a peak in the region of phase transition and mostly takes negative values.

\begin{figure}[t!]
\center
\includegraphics[width=\linewidth]{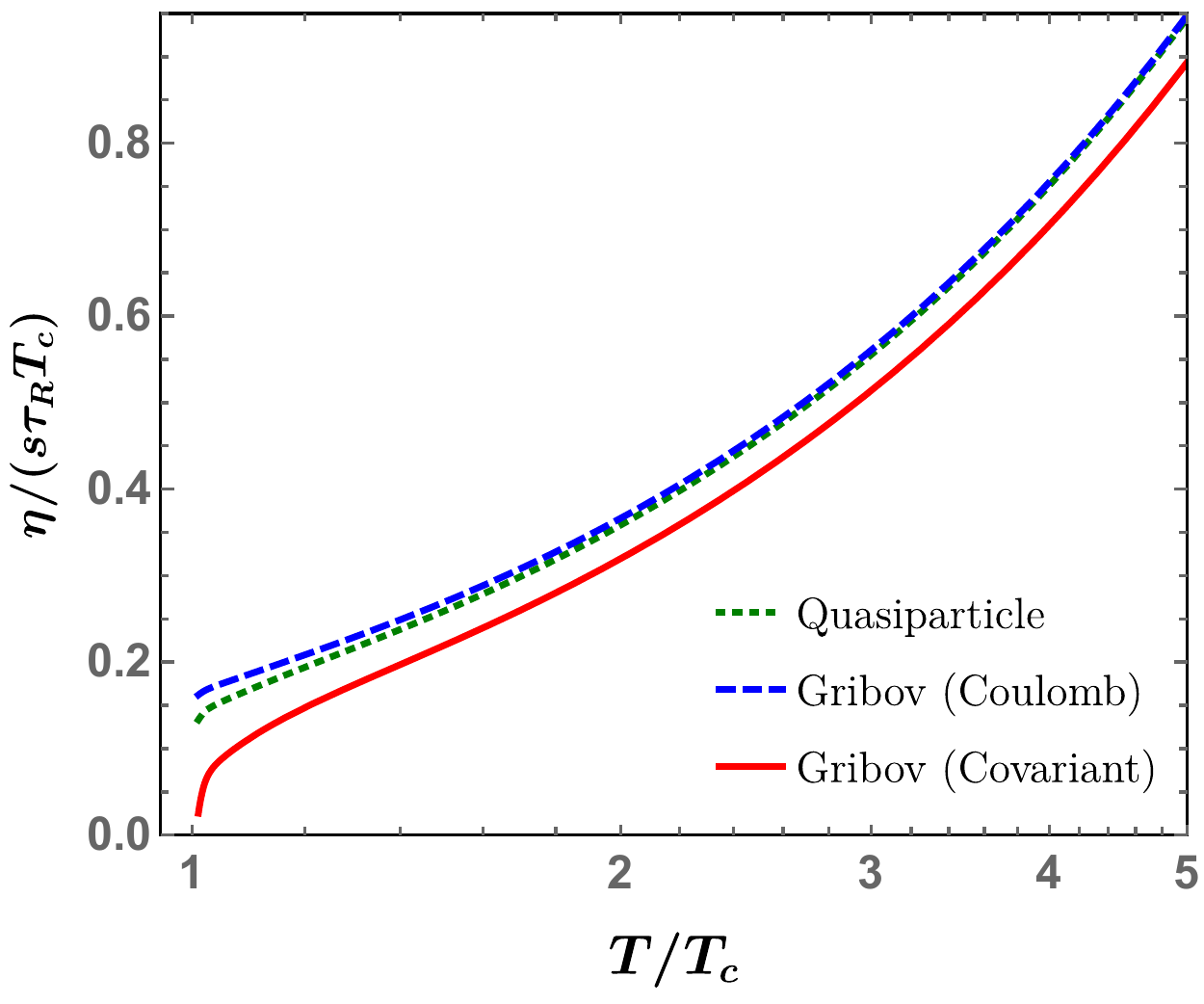}
\caption{Temperature dependence of the scaled shear viscosity coefficient of Gribov plasma obtained from Eqs.~\eqref{eta} using covariant gauge. Also shown are the results obtained using Coulomb gauge~\cite{Begun:2016lgx} and within effective mass quasiparticle model~\cite{Begun:2016lgx, Czajka:2017wdo}.}
\label{eta_comp}
\end{figure} 

In Fig.~\ref{eta_comp}, we show the temperature dependence of the scaled shear viscosity coefficient, $\eta/(s\,\tau_R\,T_c)$, of the Gribov plasma obtained from Eqs.~\eqref{eta} (red solid curve). For comparison, we also show the results obtained using Coulomb gauge with longitudinal boost invariant setup~\cite{Begun:2016lgx} (blue dashed curve) and within effective mass quasiparticle model~\cite{Begun:2016lgx, Czajka:2017wdo} (green dotted curve). We see that for all the three cases, the qualitative behavior of the scaled shear viscosity coefficient are in agreement and increases with temperature. Moreover, we observe that the results for Gribov plasma with Coulomb gauge and quasiparticle model are also in quantitative agreement. On the other hand, the scaled shear viscosity coefficient for Gribov plasma obtained in the present work using covariant kinetic theory is slightly smaller than the other two cases. This indicates that the current framework leads to stronger interaction compared to the other two cases.

\begin{figure}[t!]
\center
\includegraphics[width=\linewidth]{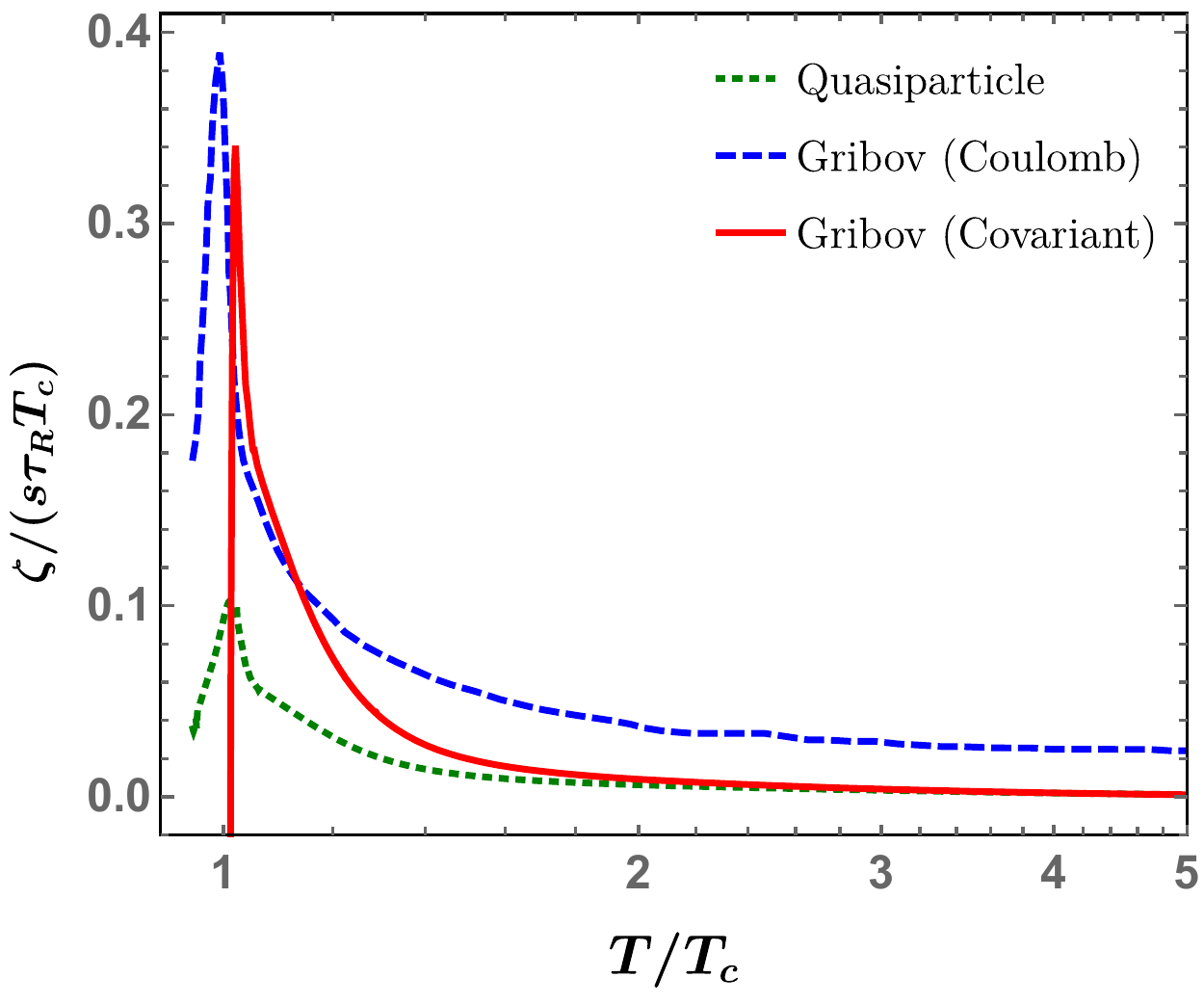}
\caption{Temperature dependence of the scaled shear viscosity coefficient of Gribov plasma obtained from Eqs.~\eqref{zeta} using covariant gauge. Also shown are the results obtained using Coulomb gauge~\cite{Begun:2016lgx} and within effective mass quasiparticle model~\cite{Czajka:2017wdo}.}
\label{zeta_comp}
\end{figure} 

In Fig.~\ref{zeta_comp}, we show the temperature dependence of the scaled bulk viscosity coefficient, $\zeta/(s\,\tau_R\,T_c)$, of the Gribov plasma obtained from Eqs.~\eqref{zeta} (red solid curve). For comparison, we also show the results obtained using Coulomb gauge with longitudinal boost invariant setup~\cite{Begun:2016lgx} (blue dashed curve) and within effective mass quasiparticle model~\cite{Czajka:2017wdo} (green dotted curve). We see that for all the three cases, the qualitative behavior of the scaled bulk viscosity coefficient are in agreement, i.e., has a peak near $T_c$ and then decreases. Moreover, we observe that the for the Gribov calculations using Coulomb gauge and covariant gauge, the peak value of scaled bulk viscosity coefficient are also in numerical agreement compare to the quasiparticle model result. However, as the temperature increases, the result obtained using covariant gauge tends to coincide with the effective mass quasiparticle calculation.


\section{Summary and outlook}

In this paper, we studied the thermodynamics and transport properties of a plasma consisting of gluons whose propagator is improved by the Gribov prescription. We first constructed the thermodynamics of Gribov plasma using the gauge invariant Gribov dispersion relation for interacting gluons. Further, we formulated, for the first time, a covariant kinetic theory for the Gribov plasma and determined the mean-field contribution in the Boltzmann equation when the Gribov parameter in the dispersion relation is considered to be temperature dependent. This resulted in a quasiparticle like framework with a bag correction to pressure and energy density. The temperature dependence of the Gribov parameter and bag pressure was fixed by matching with lattice results for a system of gluons. Finally we calculated the temperature dependence of the transport coefficients, bulk and shear viscosities and compared them with existing results for Gribov plasma using Coulomb gauge as well as within effective mass quasiparticle model.

Looking forward, it will be interesting to extend the present framework for QGP system and include quark degrees of freedom via effective mass quasiparticle model. For the QGP system, one can also generalize the present formulation to incorporate finite chemical potential. It is also straightforward to derive second-order dissipative hydrodynamics within the present framework as done in Refs.~\cite{Tinti:2016bav, Czajka:2017wdo} for temperature dependent quasiparticle masses. We leave these for future works.


\medskip

\section*{Acknowledgements}
The authors acknowledge useful discussions with Aritra Bandyopadhyay. A.J. is supported in part by the DST-INSPIRE faculty award under Grant No. DST/INSPIRE/04/ 2017/000038. N.H. is supported in part by the SERB-SRG under Grant No. SRG/2019/001680.


\end{document}